\documentclass[onecolumn, draft, 12pt]{IEEEtran}
\usepackage{amsmath,cases}
\usepackage{amsthm}
\usepackage{amssymb}
\usepackage{cite}
\usepackage{amsfonts}
\usepackage{enumerate}
\usepackage[final]{graphicx}
\usepackage{multirow}

\newcommand{\ttheta}{\ensuremath{\theta}}
\newcommand{\isrx}{\ensuremath{x_i}}
\newcommand{\ieta}{\ensuremath{n_i}}

\newcommand{\frx}{\ensuremath{y_{_{L}}}}

\newcommand{\powcnst}{\ensuremath{P_{\rm T}}}

\newcommand{\nrmfrx}{\ensuremath{z_{_{L}}}}

\newcommand{\esttheta}{\ensuremath{\widehat{\theta}}}
\newcommand{\estthetahat}{\ensuremath{\widehat{\theta}_L}}
\newcommand{\estthetahatdet}{\ensuremath{\widehat{\theta}_L}}

\newcommand{\sigeta}{\ensuremath{\sigma_n^2}}
\newcommand{\sigv}{\ensuremath{\sigma_v^2}}

\newcommand{\w}{}

\newcommand{\E}{\mathbb{E}}

\newcommand{\recdyLtheta}{\sqrt{\frac{P_{\rm T}}{L}}\sum\limits_{i=1}^{L} f(\theta + \sigma_{i}n_{i}) + v}

\newcommand{\recdsL}{\sqrt{\rho}\sum\limits_{i=1}^{L} f\left( x_{i} \right)}

\newcommand{\recdzLtheta}{\sqrt{P_{\rm T}}\frac{1}{L}\sum\limits_{i=1}^{L}f(\theta + \sigma_{i}n_{i})+\frac{v}{\sqrt{L}}}
\newcommand{\recdzL}{\frac{y_{L}}{\sqrt{L}}}

\newcommand{\strx}{\sqrt{\rho} f( x_{i})}

\newcommand{\hwIfsqtheta}{\int\limits_{-\infty}^{\infty} f^{2}(\theta+n_i ) \;  p(n_i) dn_i}
\newcommand{\hwIfdash}{ \int\limits_{-\infty}^{\infty} f^{'}(\theta+ n_i) \;  p(n_i) dn_i }

\newcommand{\sigi}{\sigma_{i}}

\newcommand{\sigsqv}{\sigma^{2}_{v}}
\newcommand{\sigsqn}{\sigma^{2}_{n}}
\newcommand{\chsnr}{\ensuremath{ \frac{\sigsqv}{\powcnst} }}

\newtheorem{thm}{Theorem}

\newtheorem{lem}{Lemma}

\begin{document}
\title{Distributed Estimation and Detection with Bounded Transmissions over Gaussian Multiple Access Channels}
\author{Sivaraman Dasarathan, Cihan Tepedelenlio\u{g}lu, \emph{Member, IEEE}
\thanks{The authors are with the School of Electrical, Computer, and Energy Engineering, Arizona
State University, Tempe, AZ 85287, USA. (Email:
\{sdasarat, cihan\}@asu.edu). This work was supported in part by the National Science Foundation under Grant NSF FRP 1231034.} } \maketitle

\begin{abstract}
A distributed inference scheme which uses bounded transmission functions over a Gaussian multiple access channel is considered. When the sensor measurements are decreasingly reliable as a function of the sensor index, the  conditions on the transmission functions under which consistent estimation and reliable detection are possible is characterized. For the distributed estimation problem, an estimation scheme that uses bounded transmission functions is proved to be strongly consistent provided that the variance of the noise samples are bounded and that the transmission function is one-to-one. The proposed estimation scheme is compared with the amplify-and-forward technique and its robustness to impulsive sensing noise distributions is highlighted. In contrast to amplify-and-forward schemes, it is also shown that bounded transmissions suffer from inconsistent estimates if the sensing noise variance goes to infinity. For the distributed detection problem, similar results are obtained by studying the deflection coefficient. Simulations corroborate our analytical results. 
\end{abstract}

\section{Introduction} \label{sec:intro_btx}
In inference-based wireless sensor networks (WSNs), low-power sensors with limited battery and peak-power capabilities transmit their observations to a fusion center (FC) for detection of events or estimation of parameters. For distributed estimation and distributed detection, much of the literature has focused on a set of orthogonal (parallel) channels between the sensors and the FC (please see \cite{magazine, vishwanath} and the references therein). The bandwidth requirements of such an orthogonal WSN scale linearly with the number of sensors. In contrast, over multiple access channels where the sensor transmissions are simultaneous and in the same frequency band, the utilized bandwidth does not depend on the number of sensors. 

Sensors may adopt either a digital or analog method for relaying the sensed information to the FC. The digital method consists of quantizing the sensed data and transmitting with digital modulation over a rate-constrained channel. In this case, the required channel bandwidth is proportional to the number of bits at the output of the quantizer which are transmitted after pulse shaping and digital modulation. The analog method consists of transmitting unquantized data by appropriately pulse shaping and amplitude or phase modulating to consume finite bandwidth. One such method is the amplify-and-forward (AF) scheme in which sensors send scaled versions of their measurements to the FC. However, using the AF technique is not a viable option for WSNs because it requires high transmission power when the values to be transmitted are large \cite{BanavarCM}. Distributed systems which employ the AF technique for transmission of the sensed data often assume that the power amplifiers used are perfectly linear over the entire range of the sensed observations. In practice, the amplifiers exhibit nonlinear behaviour  when the amplitude of the sensed data is relatively high \cite{Bobs2012,Cripps2002,Cripps2006}. Moreover, the linear transmit amplifier characteristics required for AF are often very power-inefficient \cite{Bobs2012}, requiring the study of the effect of nonlinear transmissions on performance. Wireless sensor networks have stringent power and bandwidth constraints, therefore distributed schemes which use bounded instantaneous transmit power over multiple access channels are highly desirable. 

References \cite{tepadarsh2,tepadarsh,bantep,bantep2,Mario2009,Mario2010a,Sayeed_Type} discuss distributed estimation over Gaussian multiple access channels. In \cite{tepadarsh, tepadarsh2}, a distributed estimation scheme where the sensor transmissions are phase-modulated to make constant modulus transmissions is considered. The estimator proposed in \cite{tepadarsh} is shown to be strongly consistent for any symmetric sensing noise distribution when the noise samples are i.i.d.. In \cite{bantep,bantep2}, the mean and variance of a signal embedded in noise (not necessarily Gaussian) are estimated which are then used to estimate the SNR of the signal. In \cite{tepadarsh2,tepadarsh,bantep,bantep2}, the desired constant modulus property is achieved by phase modulating the sensed data before transmission. The authors in \cite{dastep} discuss the effect of nonlinear transmissions on the convergence speed of a consensus algorithm proposed for a distributed average consensus problem. The authors in \cite{Mario2009,Mario2010,Mario2010a,Mario2012} consider the computation of a desired function of the sensor measurements by exploiting the mathematical characteristics of multiple access channels in a fusion center based wireless sensor network. In these references, they discuss different issues such as how much synchronization, channel knowledge is required for calculating various linear and nonlinear functions at the FC using the wireless multiple-access channels and study the performance of the proposed schemes.

References \cite{tepsiva, tepsiva2, Evans_Optimal, Evans_DD_MA} discuss distributed detection using constant modulus transmissions over Gaussian multiple access channels for a binary hypothesis testing problem. Inspired by the robustness of the estimation scheme in \cite{tepadarsh}, the authors in \cite{tepsiva} and \cite{tepsiva2} proposed a distributed detection scheme where the sensors transmit with constant modulus signals over a Gaussian multiple access channel. Here again, the sensors transmit with constant modulus transmissions whose phase varies linearly with the sensed data and the performance is analysed using deflection coefficient and error exponent. In \cite{Evans_Optimal} and \cite{Evans_DD_MA}, two schemes called modified amplify-and-forward (MAF) and the modified detect-and-forward (MDF) are developed  which generalize and outperform the classic amplify-and-forward (AF) and detect-and-forward (DF) approaches to distributed detection. It is shown that MAF outperforms MDF when the number of sensors is large and the opposite conclusion is true when the number of sensors is smaller. In both the DF and MDF schemes, the sensors individually take a decision by quantizing the sensed measurement and transmit the one bit information to the FC by BPSK modulation and therefore the transmit power is always constant. Bounded transmission schemes are highly desirable and practically viable for the power constrained WSNs. In addition, bounded transmissions are robust to impulsive measurements \cite{tepadarsh2,tepadarsh,bantep,bantep2} which could happen for WSNs deployed in adverse conditions. 

In this work, we are interested in studying the effect of nonlinear transmissions with general nonlinear transmission functions from the sensors to the FC in a distributed inference framework. We will contrast this with AF, especially in settings where sensing becomes decreasingly reliable. The sensors map their observations using a bounded function before transmission to constrain the transmit power and these observations are transmitted to the FC over a Gaussian multiple access channel. Our emphasis in this paper is not so much to propose a specific estimator or a detector, rather we want to focus on studying the implications of bounded transmission schemes on distributed inference in resource constrained WSNs. Moreover, this work also studies the merits and demerits of distributed schemes involving realistic, nonlinear amplifier characteristics. We  characterize the general conditions on the sensing noise statistics and the nonlinear function under which consistent estimation and reliable detection are possible. We show that if the measurement accuracy degrades progressively in the sense that the sensing noise variance goes to infinity, bounded transmission is not useful for distributed inference. On the other hand, it is shown that AF scheme does not suffer from this issue. These conclusions are drawn by studying the fundamental metrics such the asymptotic variance and the deflection coefficient. 

\section{Distributed Estimation with Bounded Transmissions} \label{sec: DE_Bounded_TX}

\subsection{System Model}\label{subsec: DE_system_model}
\begin{figure}[tb]
\begin{minipage}{1\textwidth}
\centering
\begin{center}
\includegraphics[height=9cm,width=12cm]{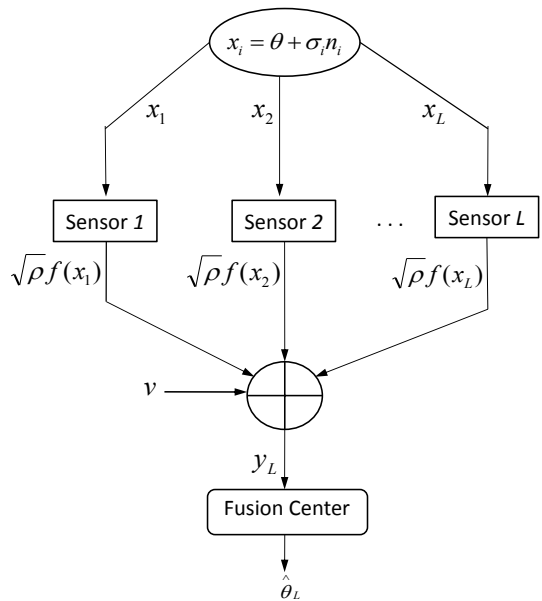}
\vspace{-0.5 in}
\caption{System Model: Bounded transmissions over Gaussian multiple access channel.}\label{fig: problem_setup}
\end{center}
\end{minipage}
\end{figure}

Consider the sensing model, with $L$ sensors,
\begin{equation}\label{eqn: ith_sensor_rx}
x_{i} = \theta + \sigma_{i}n_{i} \hspace{1 in} i = 1, \ldots, L
\end{equation}
where $\ttheta$ is an unknown real-valued parameter, $\ieta$ is symmetric real-valued noise with zero median (i.e., its probability density function (PDF) is symmetric about zero), and $\isrx$ is the measurement at the $i^{th}$ sensor. The noise samples $\ieta$ are assumed to be independent identically distributed (i.i.d.) but not necessarily with finite mean or variance. We consider a setting where the $i^{th}$ sensor transmits its measurement using a bounded function  $\strx$ over a Gaussian multiple access channel (please see Figure \ref{fig: problem_setup}) so that the received signal at the FC is given by
\begin{equation}
\label{eq:recd_signal_bt}
y_{_{L}} = \recdsL + v
\end{equation} 
where $\rho$ is a power scale factor and $v$ is the additive Gaussian noise with zero mean and variance $\sigsqv$. Parameter $\sigi$ is a deterministic scale parameter which makes the variance (when it exists) of the noise samples different for each sensor depending on how they are distributed in space and how accurate their measurements are. For instance, if the phenomenon quantified by $\theta$ happens near a sensor, it is reasonable to expect that the variances of the sensing noise would be smaller compared to those that are farther. Moreover, in case of WSNs operating in adverse conditions, the sensing noise $n_i$ could be impulsive characterized by heavy tailed distributions \cite{SignalAlpha}. We also want to point out that the received signal at the FC as modeled in \eqref{eq:recd_signal_bt} is realistic if the transmit amplifiers at the local sensors are nonlinear. 

In this paper, we study the consequences of the boundedness of $f(\cdot)$ on performance. In particular, we assume that the transmit function $f(x)$ satisfies the following conditions. \\
\textbf{Assumptions:} \\
\textbf{(A1):\;} $f(x)$ is differentiable such that $0 < f^{'}(x) \leq d$, $\forall x \in \mathbb{R}$. \\
\textbf{(A2):\;} $f(x)$ is bounded,  $\sup_{x \in \mathbb{R}} |f(x)| = c$. \\
Note that the transmitted signal at the $i^{th}$ sensor has the instantaneous power $\rho f^{2}(\w x_i)$ and it is always constrained within $\rho c^2$, which does not suffer from the problems of unbounded transmit power seen in AF schemes for which $f(x)=\alpha x$. The total transmit power from all the sensors in (\ref{eq:recd_signal_bt}) is upper bounded by $ \rho c^2 L $. We begin by considering a fixed total power constraint $\powcnst$ for the entire network implying that the per-sensor power is less than or equal to $\powcnst/L$. Clearly the per-sensor power is  a function of $L$ when $\powcnst$ is fixed.

\subsection{The Estimation Problem}\label{subsec: est_problem}
First we consider estimating $\ttheta$ from $\frx$. Let $\sigma_i$ be a deterministic sequence capturing the reliability of the $i^{\rm th}$ sensor's measurement. The received signal $\frx$ under the total power constraint is given by
\begin{equation} \label{nonlinear_model}
y_{_{L}} = \recdyLtheta .
\end{equation}
Let $z_{_{L}}$ denote the normalized received signal:
\begin{equation} \label{nonlinear_modelzl}
z_{_{L}} := \recdzL = \recdzLtheta \;,
\end{equation}
and define
$h_{\w}(\theta):= \lim_{L \rightarrow \infty} L^{-1} \sum_{i=1}^L \E_{n_{i}}\left[ f(\theta+\sigma_{i}n_{i}) \right]$ where $\E(\cdot)$ denotes expectation. We will need Kolmogorov's strong law of large numbers \cite[pp. 259]{feller} which handles the case of independent non-identically distributed RVs, due to the fact that the $\sigma_i$ are different.
\begin{thm} \label{thm_KSLLN}
Let $X_1$, $X_2$, \ldots, $X_L$ be a sequence of independent and not necessarily identically distributed RVs. Let ${\rm var}[X_k]$ denote the variance of $X_k$ and $\bar{X}_L = L^{-1} \sum_{k=1}^L X_k$ denote the partial sum of the sequence. If $\sum_{k=1}^{\infty} {\rm var}[ X_k] / {k^2} < \infty$, then,
$\bar{X}_L - \E[\bar{X}_L] \rightarrow 0$ almost surely as $L \rightarrow \infty$.
\end{thm}

Due to the law of large numbers in Theorem \ref{thm_KSLLN} we have
\begin{equation} \label{gw_theta_det}
\lim_{L \rightarrow \infty}\frac{1}{L} \sum_{i=1}^L f(\theta+\sigma_{i} n_{i} ) = h_{\w}(\theta)
\end{equation}
where we use the fact that ${\rm var}[ f(\theta+\sigma_{i} n_{i} )] \leq c^2$ are bounded. Therefore, we have $\lim_{L \rightarrow \infty} z_{_{L}} = \sqrt{\powcnst} h_{\w}(\theta) $. Due to the boundedness of $f(\cdot)$, \eqref{gw_theta_det} holds regardless of the sensing noise distributions. Consider estimating $\theta$ from, 
\begin{equation} \label{eqn: prop_estimator_det}
\estthetahatdet = h^{-1}_{\w} \left( \frac{\nrmfrx}{\sqrt{\powcnst}} \right) ,
\end{equation}
where $\nrmfrx$ is as given in \eqref{nonlinear_modelzl}. To recover $\theta$ uniquely from $h^{-1}_{\w} (\cdot)$, we need $h_{\w}(\theta)$ to be one-to-one in $\theta$ for which \textbf{(A1)} and \textbf{(A2)} are sufficient as shown in Lemma \ref{htheta_one_to_one}. 

\begin{lem} \label{htheta_one_to_one}
Let $g_{_{\sigma_i}}(\theta) := \E_{n_{i}}\left[ f(\theta+\sigma_{i}n_{i}) \right]$ and suppose that the assumptions \textbf{(A1)} and \textbf{(A2)} hold. Then, $h_{\w}(\theta)$ is one-to-one in $\theta$.
\end{lem}

\begin{IEEEproof}
Differentiating $g_{_{\sigma_i}}(\theta)$ with respect to $\theta$, we have
\begin{align} \label{eq:htheta_one_to_one}
\nonumber
g_{_{\sigma_i}}(\theta)  & = \int\limits_{-\infty}^{\infty} f(\theta+\sigma_{i} n_{i} ) p(n_{i})d n_{i} \;,\\
\frac{\partial g_{_{\sigma_i}}(\theta) } {\partial \theta}  & =  \int\limits_{-\infty}^{\infty} \frac{\partial f(\theta+\sigma_{i} n_{i} )  } {\partial \theta} p(n_{i})d n_{i} \;, \\ 
 & >0 \;,
\end{align}
where we have applied Corollary 5.9 in \cite[pp. 46]{bartle2011} using assumptions \textbf{(A1)} and \textbf{(A2)} to move the derivative inside the integral in \eqref{eq:htheta_one_to_one}. The last inequality follows from the fact that convex combination of positive valued functions is positive. Therefore, $g_{_{\sigma_i}}(\theta)$ is a strictly increasing function of $\theta$. Since $h_{\w}(\theta)$ is a convex combination of strictly increasing and differentiable functions, we have $h^{'}(\theta) >0, \theta >0$. Therefore, $h_{\w}(\theta)$ is a strictly increasing function and thus it is one-to-one in $\theta$.
\end{IEEEproof}

We now state a Lemma about a convergent sequence which will be used in the sequel.
\begin{lem} \label{cessaro_sum}
Let $a_i$ be a converging sequence such that $\lim_{i \rightarrow \infty} a_i = a$. Then, the partial sums of the sequence also converge to $a$: $\lim_{L \rightarrow \infty} L^{-1}\sum_{i=1}^L a_i=a$.
\end{lem}
\begin{IEEEproof}
Please see \cite[pp. 411]{porat1994}.
\end{IEEEproof}

An estimator $\estthetahat$ is strongly consistent if $\estthetahat$ converges to the true value $\theta$ almost surely as $L \rightarrow \infty$ \cite{porat1994}. Now we establish the strong consistency of the class of estimators $\estthetahat$ in (\ref{eqn: prop_estimator_det}) in Theorem \ref{thm2_btx}.

\begin{thm} \label{thm2_btx} 
Let the assumptions \textbf{(A1)} and \textbf{(A2)} hold. Let $\sigma_{\rm max} := \max_{i} \sigma_{i}$ be finite. Then, the estimator $\estthetahatdet$ in \eqref{eqn: prop_estimator_det} is strongly consistent.
\end{thm}

\begin{IEEEproof}
Since $f(x)$ is a bounded function by assumption \textbf{(A2)}, the variances ${\rm var}[ f((\theta+\sigma_{i} n_{i} )] \leq c^2$ are bounded so that Kolmogorov's condition $\sum_{i=1}^{\infty} {{\rm var}[ f((\theta+\sigma_{i} n_{i} )]} / {i^2} \leq \rho^2  c^2\pi^2/6$ is satisfied. Therefore the strong law of large numbers for the non-identically distributed random  variables (RVs) is applicable and $\nrmfrx \rightarrow \sqrt{\powcnst} h_{\w}(\theta)$ almost surely. Since $f^{'}(x) >0$ by assumption \textbf{(A1)}, it follows from Lemma \ref{htheta_one_to_one} that $h(\theta)$ is one-to-one in $\theta$. Due to the fact that $\estthetahatdet$ is a continuous function of $\nrmfrx$, $\estthetahatdet \rightarrow \theta$ almost surely \cite[Thm 3.14]{porat1994} proving that the estimator in \eqref{eqn: prop_estimator_det} is strongly consistent.
\end{IEEEproof}

On the other hand, if the sensing becomes increasingly unreliable, $\sigma_{i} \rightarrow \infty$ as $i \rightarrow \infty$, then the estimator in \eqref{eqn: prop_estimator_det} is not consistent and $\theta$ can not be estimated from $z_{_{L}}$. A more formal statement is presented next as a theorem.
\begin{thm} \label{thm3_btx} 
Let the assumptions \textbf{(A1)} and \textbf{(A2)} hold and $\sigma_{i}$ be a deterministic sequence such that $\sigma_{i} \rightarrow \infty$ as $i \rightarrow \infty$, then $h_{\w}(\theta)$ is independent of $\theta$.
\end{thm}
\begin{IEEEproof}
First we note that due to assumption \textbf{(A2)}, the variances ${\rm var}[ f((\theta+\sigma_{i} n_{i} )]$ are bounded. According to Kolmogorov's strong law of large numbers for non-identically distributed random variables, we have
\begin{align}
\label{eqn2:thm3_btx} 
h_{\w}(\theta)  &= \sqrt{\powcnst} \lim_{L \rightarrow \infty} \int\limits_{-\infty}^{\infty} \frac{1}{L} \sum_{i=1}^L  f(\theta+\sigma_{i} n_{i} ) p(n_{i})d n_{i}\\
\label{eqn3:thm3_btx}	
	 &= \sqrt{\powcnst} \int\limits_{-\infty}^{\infty} \lim_{L \rightarrow \infty}  \frac{1}{L} \sum_{i=1}^L  f(\theta+\sigma_{i} n_{i} ) p(n_{i})d n_{i} \\	
\nonumber
	 &= \sqrt{\powcnst} \int\limits_{-\infty}^{0} \lim_{L \rightarrow \infty}  \frac{1}{L} \sum_{i=1}^L  f(\theta+\sigma_{i} n_{i} ) p(n_{i})d n_{i} \\ \label{eqn4:thm3_btx}	
	 & \;\;\;\;\;\; + \sqrt{\powcnst} \int\limits_{0}^{\infty} \lim_{L \rightarrow \infty}  \frac{1}{L} \sum_{i=1}^L  f(\theta+\sigma_{i} n_{i} ) p(n_{i})d n_{i} \\
\label{eqn5:thm3_btx}	
	 &= \sqrt{\powcnst} \left( -\frac{c_2}{2} + \frac{c_1}{2}\right)=\frac{(c_1-c_2)}{2} 
\end{align}
for some $c_1 \leq c$, $c_2 \leq c$. We have exchanged the summation and expectation to get \eqref{eqn2:thm3_btx}. We have used assumption \textbf{(A2)} to apply bounded convergence theorem \cite[pp. 288]{bartle} to move the limit in \eqref{eqn2:thm3_btx} inside the integral as in \eqref{eqn3:thm3_btx}. In \eqref{eqn4:thm3_btx}, we have used Lemma \ref{cessaro_sum} for the sequence $f(\theta+\sigma_{i} n_{i} )$ along with the fact that $f(x)$ converges to some constant as $|x| \rightarrow \infty$ by the virtue of assumptions \textbf{(A1)} and \textbf{(A2)}. Thus if $\sigma_{i} \rightarrow \infty$, then $z_{_{L}} \rightarrow (c_1-c_2)/2$ almost surely so that $h_{\w}(\theta)$ is independent of $\theta$ and therefore $\theta$ can not be recovered from  $h_{\w}(\theta)$ and the theorem is proved. 
\end{IEEEproof}

It might seem an obvious conclusion that decreasingly reliable measurements yield inconsistent estimates. However, for AF transmissions, Theorem \ref{thm3_btx} does not hold, as will be discussed in Section \ref{sec: compAF}.

Theorem \ref{thm3_btx} indicates that if sensors use a bounded function to transmit their measurements to the FC, there is a penalty incurred when the variance of the noise samples are going to infinity. When the noise samples are very high in magnitude, the sensors will be transmitting the boundary values ($c_1$ or $-c_2$) most of the time. These boundary values do not contain any information on the quantity of interest $\theta$, therefore we can not construct any useful estimator of $\theta$ from $\nrmfrx$.


We like to point out that the assumption \textbf{(A2)} is not necessary for Theorems \ref{thm2_btx} and \ref{thm3_btx} to hold. It is sufficient if $f(x)$ is just an increasing function such that the variances ${\rm var}[ f(\theta+\sigma_{i} n_{i} )]$ are bounded and boundedness of $f(x)$ is not necessary. For instance, consider the function $f(x)={\rm sign}(x) |x|^p$ with $0<p<1/2$. This is not a bounded function, however ${\tilde{\sigma_i}}^2:={\rm var}[ {\rm sign}((\theta+\sigma_{i} n_{i} ))  |(\theta+\sigma_{i} n_{i} )|^p ]$ exists when $n_i$ is a alpha-stable random variable \cite[pp. 18]{Gennady1994} and the sequence ${\tilde{\sigma_i}}^2$ is bounded if $\sigma_{\rm max} := \max_{i} \sigma_{i}$ is finite. Therefore, Kolmogorov's strong law of large numbers is still applicable and it is possible to estimate $\theta$ from $z_{_{L}}$ in \eqref{nonlinear_modelzl}. 

\subsection{Asymptotic Normality of the Estimator}\label{sec: asympt_norm_est}

We now investigate the asymptotic normality of the estimator in (\ref{eqn: prop_estimator_det}). For the sake of simplicity we assume that $n_i$ are i.i.d. and $\sigma_i =1, i = 1, \ldots, L$. 
\begin{thm} \label{thm4_btx}
Let the assumption \textbf{(A1)} hold and suppose that $\sigma_i =1, i = 1, \ldots, L$. Let $n_i$ be i.i.d. and $v \sim \mathcal{N} (0, \sigv)$, then
$\sqrt{L}\left( \estthetahat-\ttheta \right)$ is asymptotically normal with zero mean and variance given by
\begin{equation} \label{eqn: Asymp-var}
AsV = \frac{ \hwIfsqtheta - h^{2}_{\w}(\theta) +\frac{\sigv}{\powcnst} }{ {\left(\hwIfdash\right)}^2 } .
\end{equation}
\end{thm}

\begin{IEEEproof}
Due to the central limit theorem, we see that $\sqrt{L}\left[\nrmfrx-h_{\w}(\theta)\right]$ is asymptotically normal with zero mean and  variance $\sigma^2$ given by
\begin{equation} \label{eqn: Asymp-vara}
\sigma^2  = \powcnst \left [\hwIfsqtheta - h^{2}_{\w}(\theta) \right] + \sigv.
\end{equation}
Applying  \cite[Thm 3.16]{porat1994} the asymptotic variance of the estimator in (\ref{eqn: prop_estimator_det}) is given by
\begin{equation}\label{eqn: asymp_var2}
AsV = G^2 \sigma^2 
\end{equation}
where
\begin{align}
\nonumber
G:= \frac{\partial h^{-1}_{\w}(\frac{z_{_{L}}}{\sqrt{\powcnst}} ) } {\partial z_{_{L}}}\bigg|_{{z_{_{L}}} =\sqrt{\powcnst} h_{\w}(\theta)} &=\frac{ 1} { h^{'}_{\w}\left( h^{-1}_{\w} \left (  \frac{z_{_{L}}}{\sqrt{\powcnst}}   \right) \right)} \bigg|_{{z_{_{L}}}=\sqrt{\powcnst} h_{\w}(\theta)} \\
&= \frac{1}{\sqrt{\powcnst} h^{'}_{\w}(\theta)} 
\end{align}
Substituting $G$ in (\ref{eqn: asymp_var2}) and simplifying we obtain the theorem.
\end{IEEEproof}
 
\subsection{Comparison with Amplify and Forward Scheme}\label{sec: compAF}
For the AF scheme, the transmitted signal at the $i^{th}$ sensor is given by $\alpha_L \isrx$ where $\alpha_L$ depends on
the number of sensors $L$ to ensure the total power constraint, but is independent of $\isrx$ \cite{tepadarsh}, \cite{cui2007}, \cite{gastparb2003}. To begin with, we focus on the case when $n_i$ are i.i.d., and choosing $\alpha_L$ identical across sensors. In what follows, we will show that the scheme in (\ref{eqn: prop_estimator_det}) is superior to AF when the sensing noise has a heavy-tailed density.

The received signal for the AF scheme is given by
\begin{equation}\label{eqn: fusion_rx_AF}
y_{_L} = \alpha_L \sum_{i=1}^L (\ttheta+ \sigma_i \ieta)+v\;.
\end{equation}
We have already seen that the per-sensor power $\alpha_L^2 (\ttheta+ \sigma_i \ieta)^2$ is a RV with unbounded support, when the PDF of the sensing noise has support over the entire real line. This is undesirable especially for low-power sensor networks with limited peak-power capabilities. Using a bounded transmission function is preferable to AF, with respect to the management of the instantaneous transmit power of sensors.

Since the total instantaneous power is random for AF, the total power is defined as an average $\powcnst  = \alpha_L^2 \sum_{i=1}^{L} \E [(\ttheta+ \sigma_i \ieta)^2]$, where the expectation is taken with respect to the sensing noise distribution. We will consider a total power constraint case where $\powcnst$ is not a function of $L$ so that $\alpha_L = \sqrt{\frac{\powcnst}{\sum_{i=1}^L (\theta^2+ \sigma_i^2 \sigeta)}}$ where $\sigeta$ is the variance of $n_i$. For the AF scheme the estimator is given by $\esttheta_{AF} = \frx/(L \alpha_L)$ so that
\begin{equation} \label{asvaf}
(\esttheta_{AF} - \theta) = \frac{1}{L} \sum_{i=1}^L \sigma_i \ieta + \frac{1}{L} \sqrt{\frac{\sum_{i=1}^L  (\theta^2+ \sigma_i^2 \sigeta)}{\powcnst}} \; v \;.
\end{equation} 
The normalized multiple access channel output for the AF scheme is proportional to the sample mean, which is not a good estimator of $\theta$ when the sensing noise is heavy-tailed. As a specific example, consider the case when $n_i$ is Cauchy distributed. From (\ref{asvaf}) it is clear that $(\esttheta_{AF} - \theta) \rightarrow 0$ is not possible since the sample mean  $L^{-1}\sum_{i=0}^L \sigma_i \ieta$ is Cauchy distributed for any value of $L$. Since the sample mean is not a consistent estimator for Cauchy noise, the AF approach over multiple access channels fails for such a heavy-tailed distribution. On the other hand, the estimator proposed in (\ref{eqn: prop_estimator_det}) is strongly consistent in the presence of any sensing noise distribution, including Cauchy distribution. This example illustrates that the inherent robustness of using the bounded transmission function in the presence of heavy-tailed sensing noise distributions. The sample mean, ``computed'' by the multiple access channel in the AF approach, is highly suboptimal, and sometimes not consistent like in the Cauchy case, whereas in the proposed approach the channel computes a noisy and normalized version of the function of the sensed samples, from which a consistent estimator can be constructed for any sensing noise distribution. 

We saw that bounded transmissions are more robust to impulsive sensing noise compared to AF. On the other hand, AF can be superior to bounded transmissions if the sensed data are decreasingly reliable ($\sigma_i \rightarrow \infty$). Recall Theorem \ref{thm3_btx} which says that if $\sigma_i \rightarrow \infty$, then the estimator in (\ref{eqn: prop_estimator_det}) is not consistent. It is clear from (\ref{asvaf}) that AF is strongly consistent provided that $L^{-1}\sum_{i=0}^L \sigma_i \ieta$ converges to zero. A sufficient condition for this is given by Theorem \ref{thm_KSLLN} which is given by $\sum_{i=1}^\infty \sigma_i^2/i^2 < \infty$ in this case. It is possible for $\sigma_i \rightarrow \infty$ while $\sum_{i=1}^\infty \sigma_i^2/i^2 < \infty$, when the variances of $n_i$ exist. For example, if $\sigma_i=\sqrt{i} \sigma$ for some $\sigma>0$, then $\sigma_i \rightarrow \infty$ as $i \rightarrow \infty$. However, $\sum_{i=1}^\infty \sigma_i^2/i^2 = \sigma^2 \sum_{i=1}^\infty i^{\frac{-3}{2}} < \infty$. Therefore, in this case the strong law of large numbers holds, and the AF scheme is consistent. Whereas bounded transmission schemes fail to be consistent as was proved in Theorem \ref{thm3_btx} irrespective of at what rate $\sigma_i$ goes to $\infty$. Thus, AF is consistent over a less strict set of conditions on $\sigma_i$, even though it suffers from unlimited peak power. 

\section{Distributed Detection with Bounded Transmissions} \label{sec: DD_Bounded_TX}
For the distributed estimation problem, we saw that consistency requires that $f(\cdot)$ is one-to-one. For distributed detection this is not necessary, since we do not seek to estimate $\theta$ but to distinguish between two hypothesis. Indeed, conventionally, $f(\cdot)$ is chosen as a quantizer in distributed detection. In this section, we want to address the choice of $f(\cdot)$ whether it is a quantizer, or an invertible bounded function. We also want to study the consequences of boundedness for $f(\cdot)$ through the deflection coefficient.

\subsection{System Model}\label{subsec: DD_system_model_btx}

Consider a binary hypothesis testing problem with two hypotheses $H_0$, $H_1$ where $P_0$, $P_1$ are their respective prior probabilities. Let the sensed signal at the $i^{th}$ sensor be,
\begin{equation}
\label{eq:DD_system_model_btx}
x_i =
\begin{cases}
\theta + \sigma_{i}  n_{i}& \mathrm{under \:} H_{1}\\
\sigma_{i}  n_{i}  & \mathrm{under \:} H_{0}\end{cases}
\end{equation}
$i=1, \ldots, L$, $\theta>0$ is a known parameter whose presence or absence has to be detected, $L$ is the total number of sensors in the system, and $n_i$ is the noise sample at the $i^{th}$ sensor. As explained in Section \ref{subsec: DE_system_model}, $\sigma_{i} >0$ is a deterministic scale parameter. The sensing noise samples are i.i.d, have zero median but they need not be bounded or have any finite moments. We consider a setting where the $i^{th}$ sensor transmits its measurement using a bounded function  $\strx$ over a Gaussian multiple access channel so that the received signal at the FC is given by \eqref{eq:recd_signal_bt} where $\rho$ is a power scale factor and $f(x)$ satisfies the same conditions as in Section \ref{subsec: DE_system_model}, and $v \sim  \mathcal{N}(0, \sigma^{2}_{v}) $ is the additive channel noise. Note that the power at each sensor is upper bounded by $\rho c^2$. We also assume that the total power $\rho c^2 L$ for the entire network is constrained to $P_{\rm T}$.

\subsection{The Detection Problem} \label{subsec: Detection_problem}
The received signal $y_{_{L}}$ under the total power constraint can be written as
\begin{equation}
\label{eq:recd_signal_PT_btx}
y_{_{L}} = \sqrt{\frac{\powcnst}{L}}\sum\limits_{i=1}^{L} f\left( x_{i} \right) + v \;.
\end{equation}
With the received signal in (\ref{eq:recd_signal_PT_btx}), the FC has to decide which hypothesis is true. It is well known that the optimal decision rule under the Bayesian formulation is given by:
\begin{equation}
\label{eq:opt_fusion_rule_btx}
\frac{p(y_{_{L}}|H_1)}{p(y_{_{L}}|H_0)} \overset{H_{1}}{\underset{H_{0}}{\gtrless}} \frac{P_0}{P_1}
\end{equation}
where $p(y_{_{L}}|H_i)$, is the conditional probability density function of $y_{_{L}}$ when the hypothesis $H_i$, $i \in \lbrace 0,1 \rbrace$, is true.

\subsection{Probability of Error} \label{sec: analysis_prob_error_btx}
The PDFs of $y_{_{L}}$ in \eqref{eq:opt_fusion_rule_btx} under the hypothesis $H_i$ involve $(L+1)$ convolutions and are not tractable in general. Let $P_{\rm e}$ be the probability of error at the FC:
\begin{equation}
\label{eq:prob_error_bayesian_btx}
P_{\rm e} = P_0 \Pr \left [{\rm error}|H_0 \right ] + P_1 \Pr \left [{\rm error}|H_1 \right ]
\end{equation}
where $\Pr \left [{\rm error}|H_i \right ]$ is the error probability when $H_i$ is true. Since $P_{\rm e}$ is not straightforward to evaluate, we will study a surrogate metric called the deflection coefficient (DC) \cite{picinbono, varshney, poor, kassam} to identify regimes where reliable detection is possible. The DC, depends only on the system model in \eqref{eq:recd_signal_PT_btx}, and does not depend on any detector. As we are considering a general transmission scheme at the local sensors, and $P_{\rm e}$ is not tractable, it is more insightful to study the DC.

\subsection{Deflection Coefficient and its Optimization} \label{sec: analysis_n_opt_D_w_btx}
We will now define and use the deflection coefficient which reflects the output-signal-to-noise-ratio and widely used in optimizing detectors \cite{picinbono, varshney, poor, kassam}. The DC is an SNR like quantity defined as,
\begin{equation}
\label{eq:defl_coef_def_btx}
D:=\frac{1}{L} \frac{|\E[y_{_{L}}|H_1] - \E[y_{_{L}}|H_0] |^{2}}{{\rm var}[y_{_{L}}|H_0]} .
\end{equation}

When $\sigma_{i}$ is a deterministic sequence, the DC for the system in \eqref{eq:recd_signal_PT_btx} is given by
\begin{equation}
\label{eq:defl_coef_PT_btx2} 
D_{L}=\frac {\left ( L^{-1} \displaystyle \sum_{i=1}^L \int\limits_{-\infty}^{\infty} [f(\theta+\sigma_{i} n_{i} ) -  f(\sigma_{i} n_{i} )] p(n_i) dn_i \right)^2} {  L^{-1} \displaystyle  \sum_{i=1}^L \left [ \int\limits_{-\infty}^{\infty} f^2( \sigma_{i} n_{i} ) p(n_{i})d n_{i} - \left( \int\limits_{-\infty}^{\infty} f( \sigma_{i} n_{i} ) p(n_{i})d n_{i} \right)^2 \right] + \chsnr  }.
\end{equation}
We now study the conditions on the sequence $\sigma_{i}$ for $\lim_{L \rightarrow \infty} D_{_L}=0$. When this asymptotic DC is zero, the interpretation is that reliable detection is not possible. The following result establishes that if $\sigma_{i}$ goes to infinity, the asymptotic DC is zero.

\begin{thm} \label{thm5_btx}
Let $\sigma_{i}$ be a deterministic sequence such that $\lim_{i \rightarrow \infty} \sigma_{i}=\infty$, suppose that the assumptions \textbf{(A1)} and \textbf{(A2)} hold. Then, $\lim_{L \rightarrow \infty} D_{_L}=0$.
\end{thm}

\begin{IEEEproof}
Clearly the denominator of \eqref{eq:defl_coef_PT_btx2} is bounded between $(\sigma_{v}^{2} / P_{\rm T})$ and $(c^2+\sigma_{v}^{2} / P_{\rm T})$. Therefore, it suffices to show that the numerator goes to 0 as $L \rightarrow \infty$. 
Consider
\begin{align}
\label{eqn1:thm5_btx}
\lim_{L \rightarrow \infty} D_{_{L}} &= \lim_{L \rightarrow \infty} \frac{1}{L} \sum_{i=1}^L \int\limits_{-\infty}^{\infty} [f(\theta+\sigma_{i} n_{i} )- f(\sigma_{i} n_{i})] p(n_{i})d n_{i} \\
\label{eqn2:thm5_btx}
	 &= \int\limits_{-\infty}^{\infty} \lim_{L \rightarrow \infty} \frac{1}{L} \sum_{i=1}^L [f(\theta+\sigma_{i} n_{i} )- f(\sigma_{i} n_{i})] p(n_{i})d n_{i} \\
\nonumber
	 &= \int\limits_{-\infty}^{0} \lim_{L \rightarrow \infty} \frac{1}{L} \sum_{i=1}^L [f(\theta+\sigma_{i} n_{i} )- f(\sigma_{i} n_{i})] p(n_{i})d n_{i} \\ 
\label{eqn3:thm5_btx}	
	 & \;\;\; + \int\limits_{0}^{\infty} \lim_{L \rightarrow \infty} \frac{1}{L} \sum_{i=1}^L [f(\theta+\sigma_{i} n_{i} )- f(\sigma_{i} n_{i})] p(n_{i})d n_{i} \\
\label{eqn5:thm5_btx}	
	 &= \left( -\frac{c_2}{2} + \frac{c_2}{2}\right) + \left( \frac{c_1}{2} - \frac{c_1}{2}\right)=0
\end{align}
for some $c_1 \leq c$, $c_2 \leq c$ and we have used assumption \textbf{(A2)} to apply bounded convergence theorem \cite[pp. 288]{bartle} to move the limit in \eqref{eqn1:thm5_btx} inside the integral as in \eqref{eqn2:thm5_btx}. In \eqref{eqn3:thm5_btx}, we have used Lemma \ref{cessaro_sum} for the sequences $f(\theta+\sigma_{i} n_{i} )$ and $f( \sigma_{i} n_{i} )$ along with the fact that $f(x)$ converges to some constant as $|x| \rightarrow \infty$ by the virtue of assumptions \textbf{(A1)} and \textbf{(A2)}. Thus if $\sigma_{i} \rightarrow \infty$, then $\lim_{L \rightarrow \infty} D_{_{L}}= 0$.
\end{IEEEproof}  

Theorem \ref{thm5_btx} indicates that if sensors use a bounded function to transmit their measurements to the FC, there is a penalty incurred when the variance of the noise samples are very high. When the noise samples are very high in magnitude, the sensors will be transmitting the boundary values of $f(x)$, i.e., $c_1$ or $-c_2$ most of the time. These boundary values do not contain any information about the signal $\theta$ to be detected when $H_1$ is true. Hence it is not possible to distinguish between the hypothesis $H_1$ and $H_0$ and accordingly we have the asymptotic DC equal to 0.

However, if $\sigma_{i}$ are bounded, then we can show that $\lim_{L \rightarrow \infty} D_{_L}>0$ which is done next.
\begin{thm} \label{thm6_btx}
Let $\sigma_{\rm max} := \max_{i} \sigma_{i}$ be finite and suppose that the assumptions \textbf{(A1)} and \textbf{(A2)} hold. Then, $\lim_{L \rightarrow \infty} D_{_{L}}>0$.
\end{thm}

\begin{IEEEproof}
Let $g_{_{\sigma_i}}(\theta) := \int\limits_{-\infty}^{\infty} [f(\theta+\sigma_{i} n_{i} )- f(\sigma_{i} n_{i})] p(n_{i})d n_{i}$. To show $\lim_{L \rightarrow \infty} D_{_{L}}>0$, it suffices to show that $g_{_{\sigma_i}}(\theta) >0$, $\forall \theta >0$ for some $i$. Using the assumption \textbf{(A1)} we have,
\begin{align}
\nonumber
g_{_{\sigma_i}}(\theta)  & = \int\limits_{-\infty}^{\infty} [f(\theta+\sigma_{i} n_{i} )- f(\sigma_{i} n_{i})] p(n_{i})d n_{i} \;,\\  \label{eq:one_to_one}
\frac{\partial g_{_{\sigma_i}}(\theta) } {\partial \theta}  & =  \int\limits_{-\infty}^{\infty} \frac{\partial f(\theta+\sigma_{i} n_{i} )  } {\partial \theta} p(n_{i})d n_{i} \;, \\ 
 & >0 \;,
\end{align}
where we have applied Corollary 5.9 in \cite[pp. 46]{bartle2011} using assumptions \textbf{(A1)} and \textbf{(A2)} to move the derivative in
\eqref{eq:one_to_one} inside the integral. The last inequality follows from the fact that convex combination of positive valued functions is positive. Therefore, $g_{_{\sigma_i}}(\theta)$ is strictly an increasing function of $\theta$. When $\theta =0$, clearly $g_{_{\sigma_i}}(0)=0$ and together with the fact that  ${\partial g_{_{\sigma_i}}(\theta)} / {\partial \theta} >0$, $\forall \theta >0$, we have $g_{_{\sigma_i}}(\theta) >0$, $\forall \theta >0$.
\end{IEEEproof}

Theorem \ref{thm6_btx} says that if the deterministic $\sigma_{i}$ are bounded, then the asymptotic DC is positive which means that reliable detection is possible in this regime. 

Next we will prove that for the DC to be greater than zero, we do not need $f(x)$ to be a differentiable or strictly increasing. In the following theorem we prove that $D_{_{L}}>0$ for a uniform quantizer with bounded number of quantization levels.

\begin{thm} \label{thm8_btx}
Let $\sigma_{\rm max} := \max_{i} \sigma_{i}$ be finite and suppose that $f(x)$ is a uniform quantizer with $M$ levels such that
\begin{equation}
\label{eq:quantizer_function}
f(x) =
\begin{cases}
k \Delta \;, &  (k-\frac{1}{2}) \Delta \leq x < (k+\frac{1}{2}) \Delta  \;,\\
K \Delta \;, &   x \geq (K+\frac{1}{2}) \Delta  \;,\\
-K \Delta \;, &  x \leq -(K+\frac{1}{2}) \Delta \end{cases}
\end{equation}
where $k=-K, -(K-1), \ldots, 0, \ldots, (K-1), K$, $M=2K+1$, $\Delta = 2 x_{\rm max} / M$ and $x_{\rm max}$ is the saturation point of the finite level quantizer. Suppose that $n_i$ has infinite support. Then, $D_{_{L}}>0$.
\end{thm}

\begin{IEEEproof}
Let $g_{_{\sigma_i}}(\theta) := \int\limits_{-\infty}^{\infty} [f(\theta+\sigma_{i} n_{i} )- f(\sigma_{i} n_{i})] p(n_{i})d n_{i}$. To show $D_{_{L}}>0$, it suffices to show that $g_{_{\sigma_i}}(\theta) >0$, $\forall \theta >0$. Note that the function $f(x)$ in \eqref{eq:quantizer_function} is non-decreasing, i.e., $f(x)-f(y) \geq 0, \forall x \geq y$. Consider
\begin{align} 
\label{eq:quantizer_function2} 
g_{_{\sigma_i}}(\theta) & = \int\limits_{-\infty}^{\infty} [f(\theta+\sigma_{i} n_{i} )- f(\sigma_{i} n_{i})] p(n_{i})d n_{i} \\ \label{eq:quantizer_function2a} 
                        & = \frac{1}{\sigma_i} \int\limits_{-\infty}^{\infty} [f(\theta+ v_{i} )- f( v_{i})] p(v_{i})d v_{i} \\ \nonumber
	 &  = \frac{1}{\sigma_i} \int\limits_{-\infty}^{-[(K+\frac{1}{2})\Delta +\theta]} [f(\theta+ v_{i} )- f( v_{i})] p(v_{i})d v_{i} \\ \nonumber
 	 &  \hspace{0.25 in} + \frac{1}{\sigma_i} \int\limits_{-[(K+\frac{1}{2})\Delta +\theta]}^{(K+\frac{1}{2})\Delta} [f(\theta+ v_{i} )- f( v_{i})] p(v_{i})d v_{i}\\
\label{eq:quantizer_function3}
 	 &  \hspace{0.25 in} + \frac{1}{\sigma_i} \int\limits_{(K+\frac{1}{2})\Delta}^{\infty} [f(\theta+ v_{i} )- f( v_{i})] p(v_{i})d v_{i} \\
 	 & \geq  \frac{1}{\sigma_i} \int\limits_{-[(K+\frac{1}{2})\Delta +\theta]}^{(K+\frac{1}{2})\Delta} [f(\theta+ v_{i} )- f( v_{i})] p(v_{i})d v_{i} \\
 	 & =  \frac{1}{\sigma_i} \displaystyle  \sum_{k=-K}^K  \int\limits_{[(k-\frac{1}{2})\Delta -\theta]}^{(k+\frac{1}{2})\Delta} \Delta p(v_{i})d v_{i} \\
\label{eq:quantizer_function4}
 	 & >  0
\end{align}
where in \eqref{eq:quantizer_function2} we substituted $v_i=\sigma_i n_i$ to get \eqref{eq:quantizer_function2a}. The inequality in \eqref{eq:quantizer_function4} follows from the fact that $\Delta >0$ and $v_i$ has infinite support (since $n_i$ has infinite support so that $v_i=\sigma_i n_i$ has infinite support as well). When $\theta =0$, clearly $g_{_{\sigma_i}}(0)=0$ and therefore, we have $D_{_{L}}>0$, $\forall \theta >0$.
\end{IEEEproof}

Theorem \ref{thm8_btx} can in fact be proved for non-uniform quantizer as long as $M \geq 2$ and $n_i$ has infinite support. 

We would ideally like to find the $f(x)$ that maximizes the DC in \eqref{eq:defl_coef_PT_btx2} but this is not tractable. However, when $\theta$ is small, and channel noise is negligible, we have a closed form expression for $f(x)$ through the locally optimal detection strategy. We now briefly discuss the use of nonlinear functions in the context of locally optimal detection. 

\subsection{Locally Optimal Detection} \label{sec: locally_optimal}
A detector is said to be locally optimal (most powerful) if it is better than any other detector in the sense of minimizing the probability of error for very small values of $\theta$ \cite{SignalAlpha}. The problem of designing optimum detectors in the presence of additive noise has a long history in the statistical signal processing literature \cite{SignalAlpha}. Usually the sensing noise corrupting the signal is assumed to be Gaussian. However there are situations when the noise is impulsive \cite{SignalAlpha}. In such scenarios, linear detector is not necessarily optimal, and therefore nonlinear functions are applied on the sensed observations to minimize the impact of impulsive sensing noise distributions with heavy tails.

In \cite{SignalAlpha}, it is shown that for a given sensing noise distribution $p(n)$, the nonlinear function $f(x)$ that would be locally optimal is given by
\begin{equation}\label{eqn: locally_optimal}
f(x)=- \frac{p^{'}(x)}{p(x)} .
\end{equation}
One may be interested in the inverse problem that given a nonlinear function $f(x)$, for which sensing noise distribution, it would be locally optimal. From \eqref{eqn: locally_optimal} it is easy to answer this question. We have,
\begin{equation}\label{eqn: noise_optimal}
p(x)=  C e^{ - \int\limits_{-\infty}^{x} f(y) dy} .
\end{equation}
Here the $p(x)$ obtained from \eqref{eqn: noise_optimal} should be a valid PDF satisfying $p(x) \geq 0$ and $\int\limits_{-\infty}^{\infty} p(x) dx=1$. For example, if $f(x)=\tanh(x)$, we get $p(x)= \pi  {\rm sech}(x) = 2 \pi e^{-x} / (1+e^{-2x})$. The ${\rm sech}(x)$ distribution behaves like the heavy-tailed Laplacian distribution when $x$ is relatively high. It is interesting to note that $\tanh(x)$ behaves like the hard clipper non-linearity \cite{SignalAlpha} which is  a bounded function and is locally optimal for Laplacian noise distribution. In fact, a closer look at \eqref{eqn: locally_optimal} reveals that if $p(x)$ behaves like an exponential density (for relatively large $x$), then the $f(x)$ that would be locally optimal would behave like a constant (for relatively large $x$). This shows that the family of increasing bounded functions are locally optimal for the family of heavy tailed sensing noise distributions. When $n$ is Gaussian, bounded $f(x)$ is no longer optimal as it is well known that $f(x)=x$ is optimal for Gaussian sensing noise. We will illustrate this in the Simulations section.

\section{Simulations} \label{sec: simulations_btx}
In this section, we corroborate our analytical results through Monte Carlo simulations for both the distributed estimation and distributed detection problems. In all of the simulations we have assumed $\sigma_i =1, i = 1, \ldots, L$.

\subsection{Distributed Estimation Performance} \label{subsec: DE_perf}

\begin{figure}[tb]
\begin{minipage}{1\textwidth}
\centering
\begin{center}
\includegraphics[height=9.5cm,width=12cm]{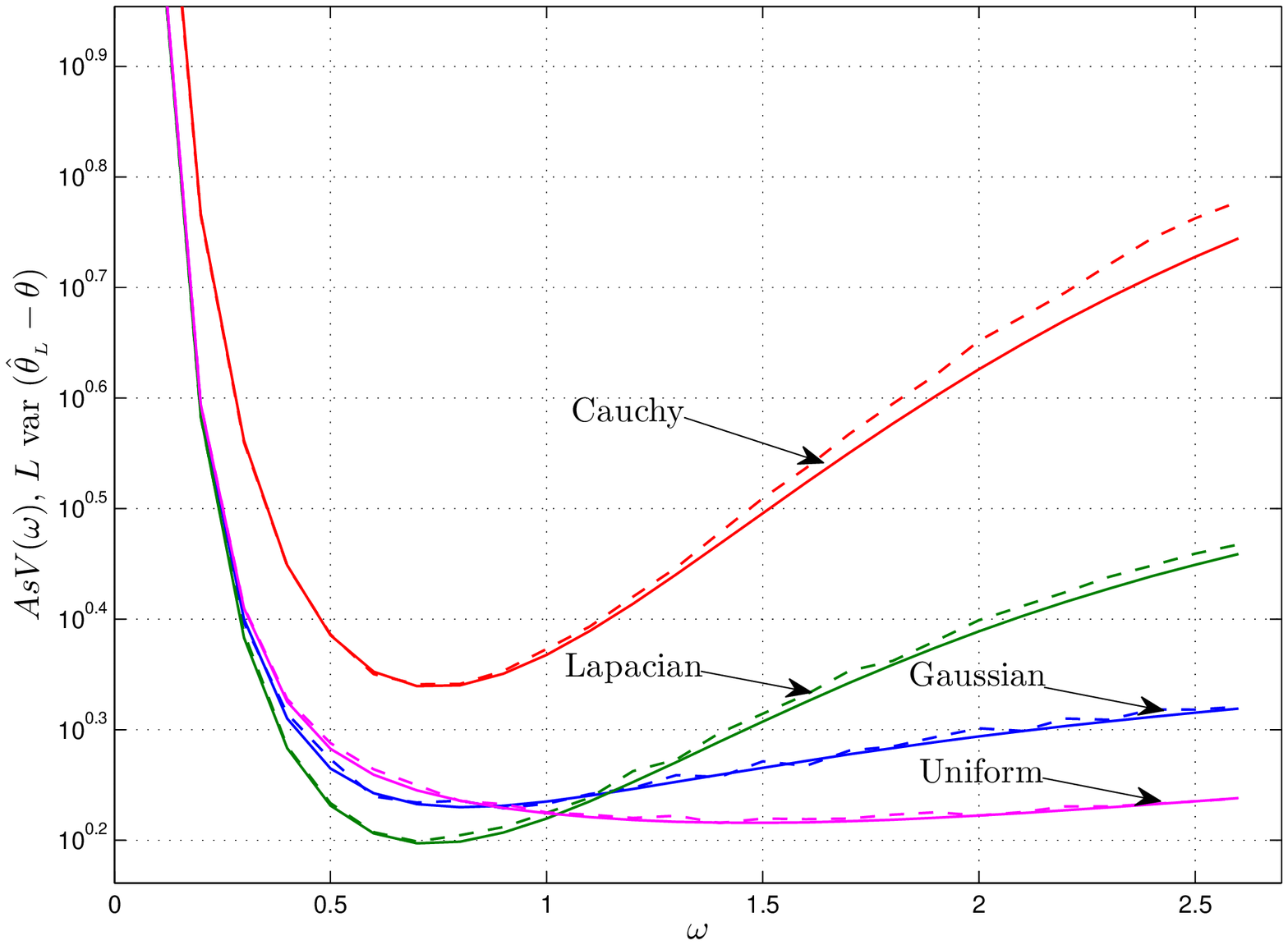}
\caption{Total Power Constraint: $f(x)=\tanh(\omega x)$, $\sigsqn$=1, $\sigsqv$=1, $\powcnst$=10,  $L$=500}\label{fig: AsV_Sim_Theory_All_Compare}
\end{center}
\end{minipage}
\end{figure}

\begin{figure}[tb]
\begin{minipage}{1\textwidth}
\centering
\begin{center}
\includegraphics[height=9.5cm,width=12cm]{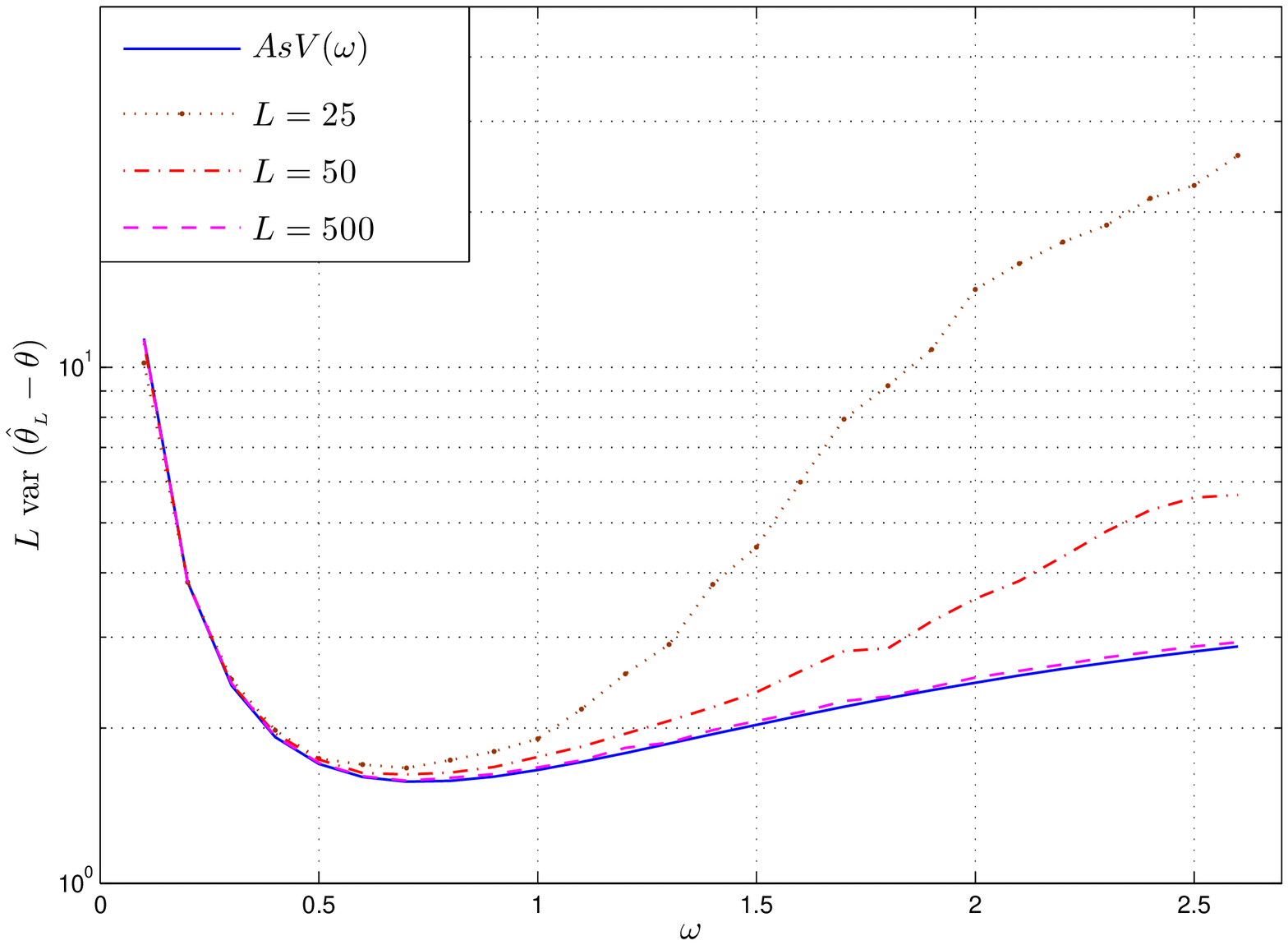}
\caption{Total Power Constraint, $n_i$ Laplacian: $f(x)=\tanh(\omega x)$, $\sigsqn$=1, $\sigsqv$=1, $\powcnst$=10,  $L$=25, 50, 500}\label{fig: AsV_Sim_Theory_Laplacian_vs_L}
\end{center}
\end{minipage}
\end{figure}

\begin{figure}[tb]
\begin{minipage}{1\textwidth}
\centering
\begin{center}
\includegraphics[height=9.5cm,width=12cm]{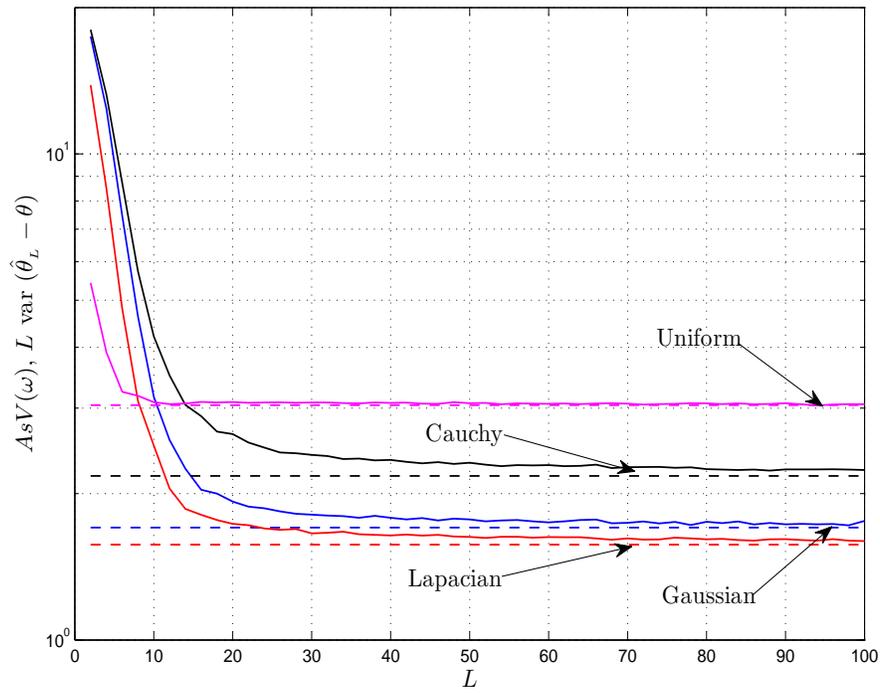}
\caption{Total Power Constraint: $f(x)=\tanh(\omega x)$, $\sigsqn$=1, $\sigsqv$=1, $\rho$=1, $\omega$=0.75}\label{fig: AsV_Sim_Theory_L_All_Compare}
\end{center}
\end{minipage}
\end{figure}

In Figure \ref{fig: AsV_Sim_Theory_All_Compare} we chose $f(x)=\tanh(\omega x)$, $\omega>0$ is a scale parameter. Here we compare $AsV(\omega)$ and $L$var(${\hat{\theta}}_{_L}-\theta)$ versus $\omega$ under the total power constraint for various distributions on the sensing noise $n_i$. We observe that the variance of the asymptotic distribution, $AsV(\omega)$ and the normalized limiting variance $L$var(${\hat{\theta}}_{_L}-\theta)$ are closer to each other when $L$ is sufficiently large. However if $L$ is smaller, we see that there is significant difference between $AsV(\omega)$ and $L$var(${\hat{\theta}}_{_L}-\theta)$ as illustrated in Figure \ref{fig: AsV_Sim_Theory_Laplacian_vs_L}. This is due to the finite sample effect, and when $L$ is increased, $L$var(${\hat{\theta}}_{_L}-\theta)$ decreases to converge its limiting value of $AsV(\omega)$. In Figure \ref{fig: AsV_Sim_Theory_L_All_Compare}, we compare $AsV(\omega)$ and $L$var(${\hat{\theta}}_{_L}-\theta)$ versus $L$. Clearly in all cases, as $L$ increases the $L$var(${\hat{\theta}}_{_L}-\theta)$ approaches its limiting values of $AsV(\omega)$. 

\begin{figure}[tb]
\begin{minipage}{1\textwidth}
\centering
\begin{center}
\includegraphics[height=9.5cm,width=12cm]{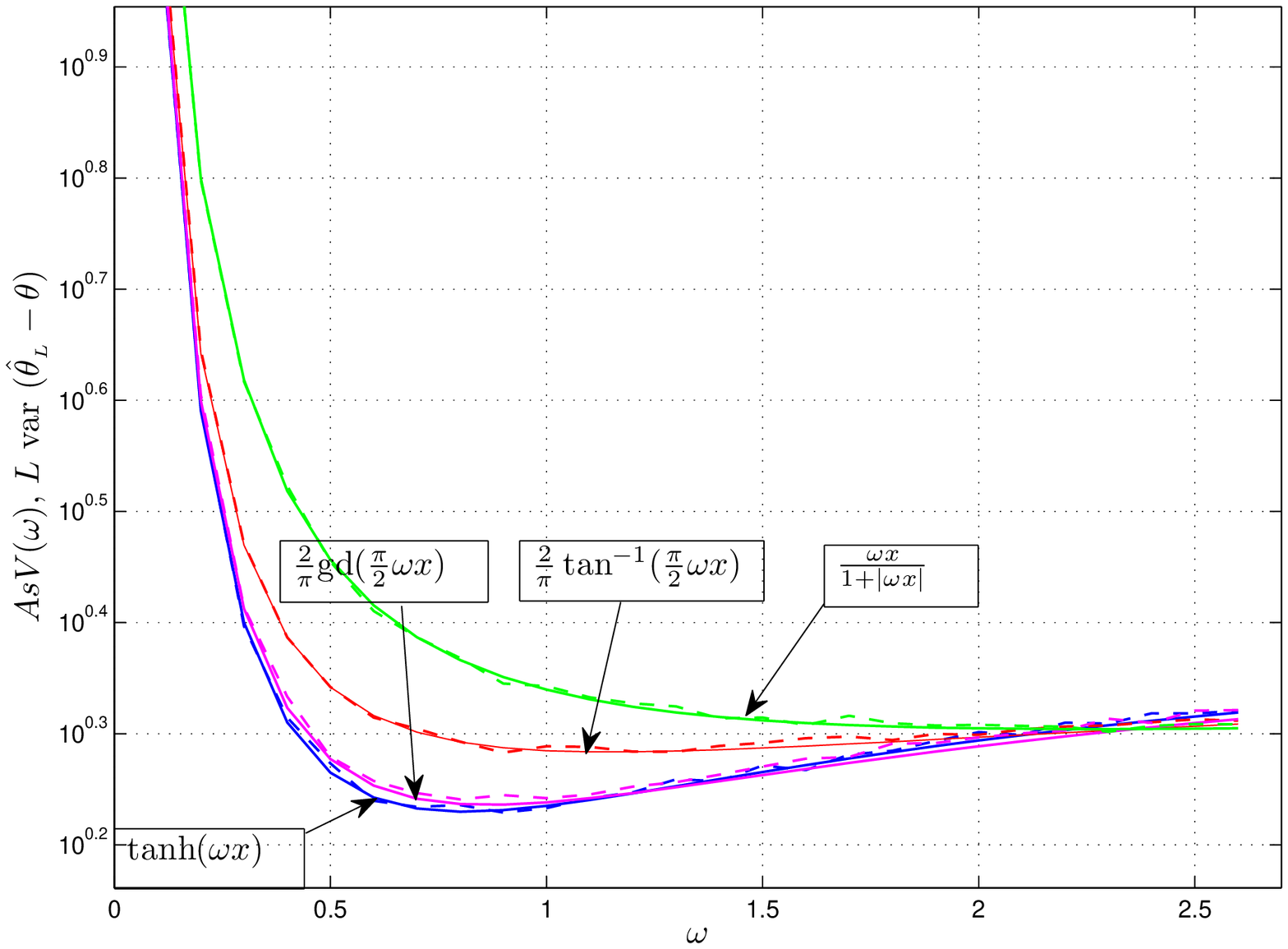}
\caption{Total Power Constraint, Different bounded functions: $\sigsqn$=1, $\sigsqv$=1, $\powcnst$=10,  $L$=500}\label{fig: AsV_w_Different_Functions}
\end{center}
\end{minipage}
\end{figure}

In Figure \ref{fig: AsV_w_Different_Functions}, we compare the performance among different bounded transmission functions when $n_i$ is Gaussian. All the functions used in this plot are appropriately normalized so that $-1 \leq f(x) \leq 1$. Here ${\rm gd}(x):= \arctan (\sinh(\omega x))$. We note that $\tanh(\omega x)$ has the lowest asymptotic variance compared to other functions. Intuitively, this is due to the fact that for a given $\omega$, $\tanh(\omega x)$ is closest to the linear function among the other functions considered here. For the Gaussian sensing noise, since linear estimator is optimal, $\tanh(\omega x)$ performs better than other functions.

\subsection{Distributed Detection Performance} \label{subsec: DD_perf}
We define the sensing and channel SNRs as $\rho_s:= \theta^2 / \sigma_{n}^{2}$, $\rho_c:={P_{\rm T}}/{\sigma_{v}^{2}}$ and assume $P_1=P_0=0.5$. Note also that $\rho=P_{\rm T}/L$ is the power at each sensor as defined in Section \ref{subsec: DE_system_model}. We used a quadratic detector based on the assumption that $y_{_L}$ in \eqref{eq:recd_signal_PT_btx} is Gaussian under both hypotheses in the simulations provided here.

\begin{figure}[tb]
\begin{minipage}{1\textwidth}
\centering
\begin{center}
\includegraphics[height=9.5cm,width=12cm]{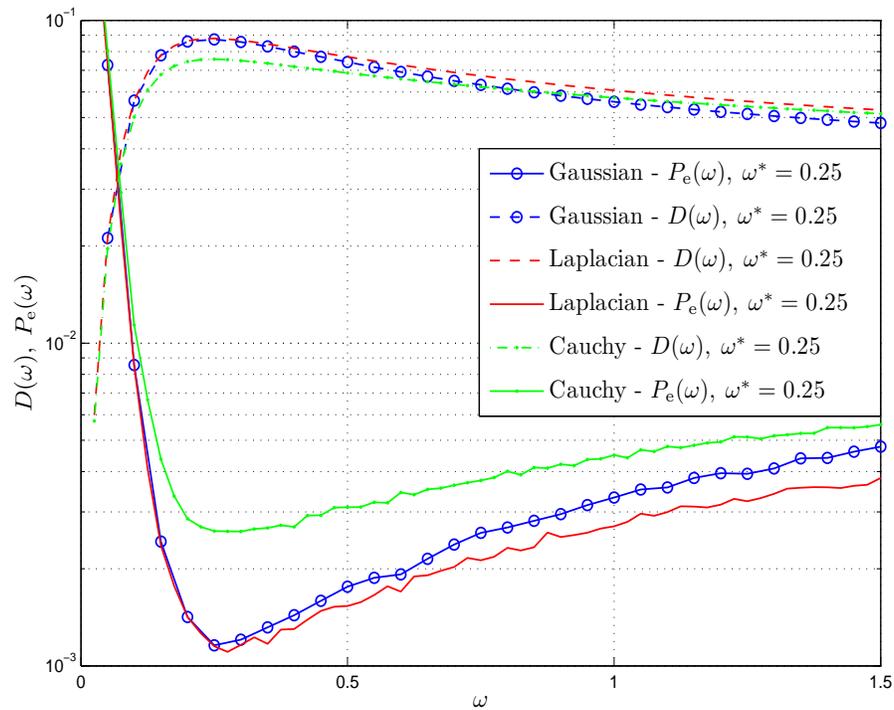}
\caption{Total Power Constraint, $f(x)=\tanh(\omega x)$, $D (\omega)\; P_{\rm e} (\omega)$ versus $\omega$, $\rho_s=10$ dB, $\rho_c=3$ dB, $L$=20}\label{fig: DC_Pe_vs_w}
\end{center}
\end{minipage}
\end{figure}

In Figure \ref{fig: DC_Pe_vs_w}, we chose $f(x)=\tanh(\omega x)$, $\omega>0$ is a scale parameter and show that maximizing the DC with respect to $\omega$ approximately results in minimizing the probability of error. Figure \ref{fig: DC_Pe_vs_w} shows the plots of $D(\omega)$ and $P_{\rm e}(\omega)$ vs $\omega$ for Gaussian, Laplacian and Cauchy sensing noise distributions where the $P_{\rm e}(\omega)$ plot is obtained using Monte-Carlo simulations. The different $\omega^{*}$ values in Figure \ref{fig: DC_Pe_vs_w} correspond to the best $\omega$ values obtained by optimizing $D(\omega)$ and $P_{\rm e}(\omega)$ respectively. It is interesting to see that the $\omega^{*}$ that minimizes $P_{\rm e}(\omega)$ is very close to that which maximizes $D(\omega)$ and thus DC is justified as a performance metric. 

\begin{figure}[tb]
\begin{minipage}{1\textwidth}
\centering
\begin{center}
\includegraphics[height=9.5cm,width=12cm]{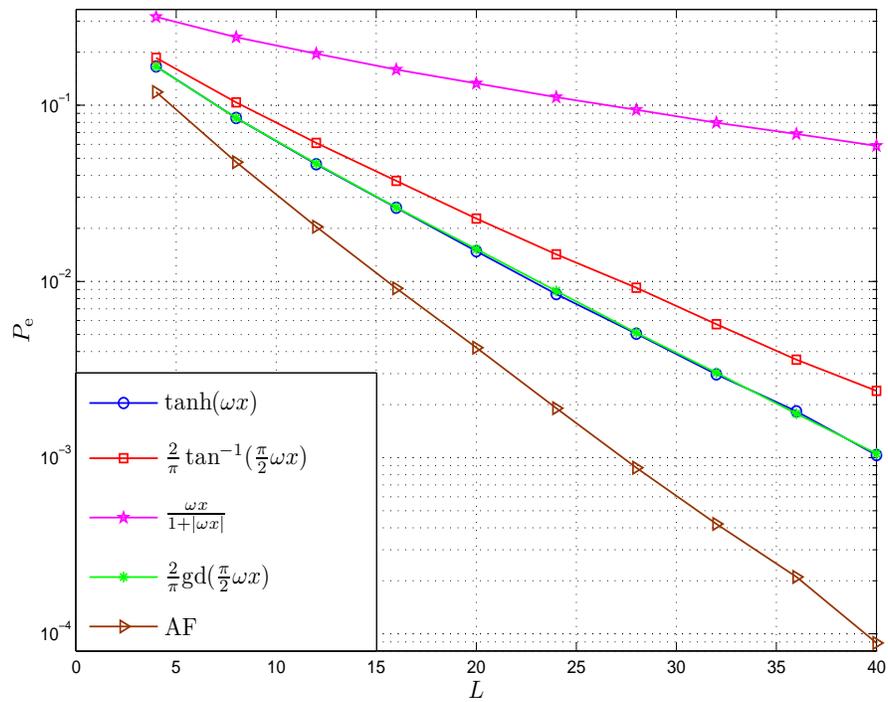}
\caption{Total Power Constraint, $n_i$ Gaussian, $P_{\rm e} $ versus $L$, $\rho_s=10$ dB, $\rho_c=0$ dB}\label{fig: Pe_vs_L_Diff_Fx}
\end{center}
\end{minipage}
\end{figure}

Finally we depict the $P_{\rm e}$ performance versus $L$ for different bounded functions in Figure \ref{fig: Pe_vs_L_Diff_Fx}. In each of these cases, $\omega^{*}$ that maximized the deflection coefficient were used. We note that AF outperforms all other functions since for the AF scheme, the detector is a linear function of observations which is optimal when $n_i$ is Gaussian. The function $\omega x / (1+|\omega x|)$ exhibits the worst performance as it has the largest deviation from the linear function compared to the other candidate functions considered in this simulation.

\section{Conclusions} \label{Conclusions}
A distributed inference scheme relying on bounded transmissions from the sensors is considered over Gaussian multiple access channels. The instantaneous transmit power is always constrained to be bounded irrespective of the random sensing noise, which is a desirable feature for low-power sensors with limited peak power capabilities. For the distributed estimation problem, the estimation scheme using bounded transmissions is shown to be strongly consistent provided that $\sigma_i$ is a bounded sequence and that the transmission function is one-to-one. For sensing noise distributions for which the sample mean is highly suboptimal or inconsistent, the proposed estimator is shown to be consistent. For heavy-tailed distributions with infinite variance like Cauchy, it is shown that the AF scheme fails, and that the bounded transmission approach is superior to AF. As long as the variance of the noise samples grow to infinity slower than linearly, AF scheme is consistent, whereas the proposed scheme fails when the variance of the noise samples go to infinity at any rate. For the distributed detection problem, it is shown that using bounded transmissions, reliable detection is possible if $\sigma_i$ is a bounded sequence. It is also shown that using bounded transmissions, reliable detection is impossible if the variance of the noise samples grow to infinity. Monte Carlo simulations are presented to illustrate the performance of several bounded transmission functions for a variety of sensing noise distributions. 

\bibliographystyle{IEEEtran}
\bibliography{boundedtx}

\newpage

%
%
%

\end{document}